\documentclass{article}
\usepackage[T1]{fontenc}
\usepackage{bookman}
\usepackage{geometry}
\usepackage{graphics}

\makeatletter

\newcommand{\LyX}{L\kern-.1667em\lower.25em\hbox{Y}\kern-.125emX\@}

\def\fnum@table{\tablename~{\bf\thetable}}
\def\fnum@figure{\figurename~{\bf\thefigure}}
\def\tablename{\footnotesize{\bf Table}}
\def\figurename{\footnotesize{\bf Figure}}

\makeatother

\begin{document}

\title{\textbf{\Huge Self-Consistency Requirement in High-Energy\protect\( \: \protect \)Nuclear\protect\( \: \protect \)Scattering}\\
\Huge }

\author{\textbf{M. Hladik\protect\( ^{1,5}\protect \), H.J. Drescher\protect\( ^{1,4}\protect \),
S. Ostapchenko\protect\( ^{1,2,3}\protect \), T. Pierog\protect\( ^{1}\protect \),
}\\
\textbf{and K. Werner}\protect\( ^{1}\protect \)\\
\\
\\
 \textit{\protect\( ^{1}\protect \)} \textit{\small SUBATECH, Université de
Nantes -- IN2P3/CNRS -- EMN,  Nantes, France }\\
\textit{\small \protect\( ^{2}\protect \) Moscow State University, Institute
of Nuclear Physics, Moscow, Russia}\\
 \textit{\small \protect\( ^{3}\protect \) Institute f. Kernphysik, Forschungszentrum
Karlsruhe, Karlsruhe, Germany}\\
\textit{\small \protect\( ^{4}\protect \) Physics Department, New York University,
New York, USA} \\
 \textit{\small \protect\( ^{5}\protect \) now at SAP AG, Berlin, Germany}\small }

\maketitle
\begin{abstract}
Practically all serious calculations of exclusive particle production in ultra-relativistic
nuclear or hadronic interactions are performed in the framework of Gribov-Regge
theory or the eikonalized parton model scheme. It is the purpose of this paper
to point out serious inconsistencies in the above-mentioned approaches. We will
demonstrate that requiring theoretical self-consistency reduces the freedom
in modeling high energy nuclear scattering enormously. 

We will introduce a fully self-consistent formulation of the multiple-scattering
scheme in the framework of a Gribov-Regge type effective theory. In addition,
we develop new computational techniques which allow for the first time a satisfactory
solution of the problem in the sense that calculations of observable quantities
can be done strictly within a self-consistent formalism.
\end{abstract}

\section{Open Problems}

With the start of the RHIC program to investigate nucleus-nucleus collisions
at very high energies, there is an increasing need of computational tools in
order to provide a clear interpretation of the data. The situation is not satisfactory
in the sense that there exists a nice theory (QCD) but we are not able to treat
nuclear collisions strictly within this framework, and on the other hand there
are simple models, which can be applied easily but which have no solid theoretical
basis. A good compromise is provided by effective theories, which are not derived
from first principles, but which are nevertheless self-consistent and calculable.
A candidate seems to be the Gribov-Regge approach, and -- being formally quite
similar -- the eikonalized parton model. Here, however, some inconsistencies
occur, which we are going to discuss in the following, before we provide a solution
to the problem.

Gribov-Regge theory \cite{gri68,gri69} is by construction a multiple scattering
theory. The elementary interactions are realized by complex objects called ``Pomerons'',
who's precise nature is not known, and which are therefore simply parameterized,
with a couple of parameters to be determined by experiment \cite{dre00}. Even
in hadron-hadron scattering, several of these Pomerons are exchanged in parallel
(the cross section for exchanging a given number of Pomerons is called''topological
cross section''). Simple formulas can be derived for the (topological) cross
sections, expressed in terms of the Pomeron parameters.

In order to calculate exclusive particle production, one needs to know how to
share the energy between the individual elementary interactions in case of multiple
scattering. We do not want to discuss the different recipes used to do the energy
sharing (in particular in Monte Carlo applications). The point is, whatever
procedure is used, this is not taken into account in the calculation of cross
sections discussed above \cite{bra90},\cite{abr92}. So, actually, one is using
two different models for cross section calculations and for treating particle
production. Taking energy conservation into account in exactly the same way
will modify the (topological) cross section results considerably. 

Another very unpleasant and unsatisfactory feature of most ``recipes'' for
particle production is the fact, that the second Pomeron and the subsequent
ones are treated differently than the first one, although in the above-mentioned
formula for the cross section all Pomerons are considered to be identical.

Being another popular approach, the parton model \cite{sjo87} amounts to presenting
the partons of projectile and target by momentum distribution functions, \( f_{i} \)
and \( f_{j} \), and calculating inclusive cross sections for the production
of parton jets as a convolution of these distribution functions with the elementary
parton-parton cross section \( d\hat{\sigma }_{ij}/dp_{\perp }^{2} \), where
\( i,j \) represent parton flavors. This simple factorization formula is the
result of cancelations of complicated diagrams and hides therefore the complicated
multiple scattering structure of the reaction, which is finally recovered via
some unitarization procedure. The latter one makes the approach formally equivalent
to the Gribov-Regge one and one therefore encounters the same conceptual problems
(see above).

\section{A Solution: Parton-based Gribov-Regge Theory}

As a solution of the above-mentioned problems, we present a new approach which
we call ``\textbf{Parton-based Gribov-Regge Theory}'': we have a consistent
treatment for calculating (topological) cross sections and particle production
considering energy conservation in both cases; in addition, we introduce hard
processes in a natural way. 

The basic guideline of our approach is theoretical consistency. We cannot derive
everything from first principles, but we use rigorously the language of field
theory to make sure not to violate basic laws of physics, which is easily done
in more phenomenological treatments (see discussion above).

\begin{figure}[htb]
{\par\centering \resizebox*{!}{0.125\textheight}{\includegraphics{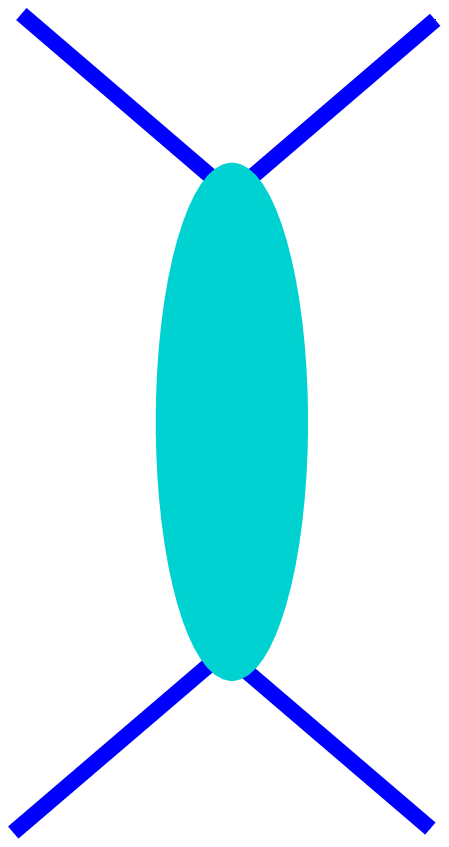}}  \( \qquad  \)\resizebox*{!}{0.125\textheight}{\includegraphics{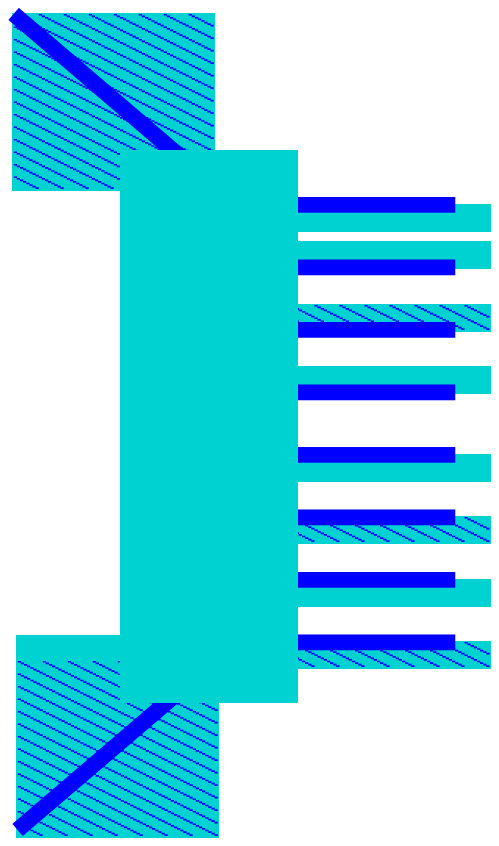}}  \par}

\caption{An elastic scattering amplitude \protect\( T_{2\rightarrow 2}\protect \) (left)
and an inelastic amplitude \protect\( T_{2\rightarrow X}\protect \) (right).\label{t}}
\end{figure}

\begin{figure}[htb]
{\par\centering \resizebox*{!}{0.12\textheight}{\includegraphics{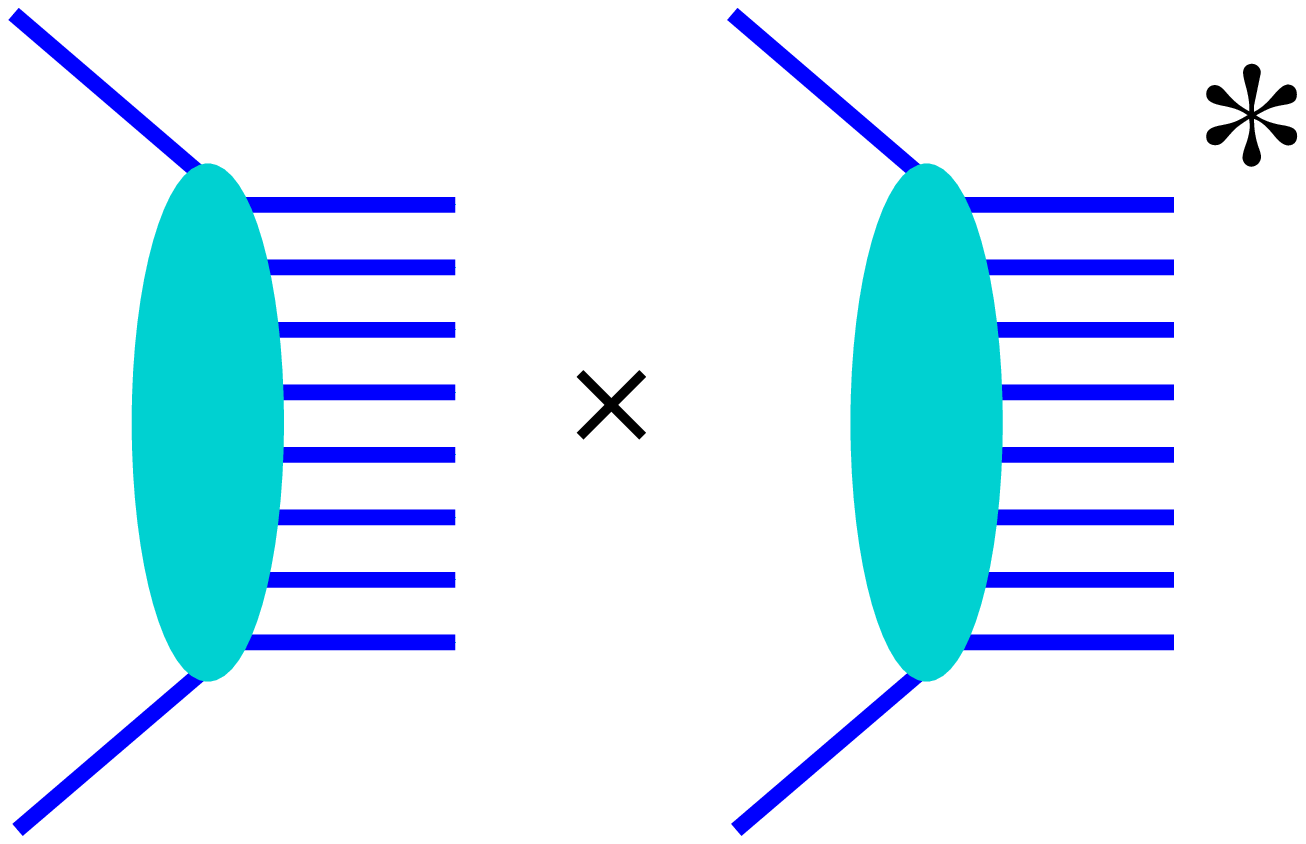}}  \resizebox*{!}{0.12\textheight}{\includegraphics{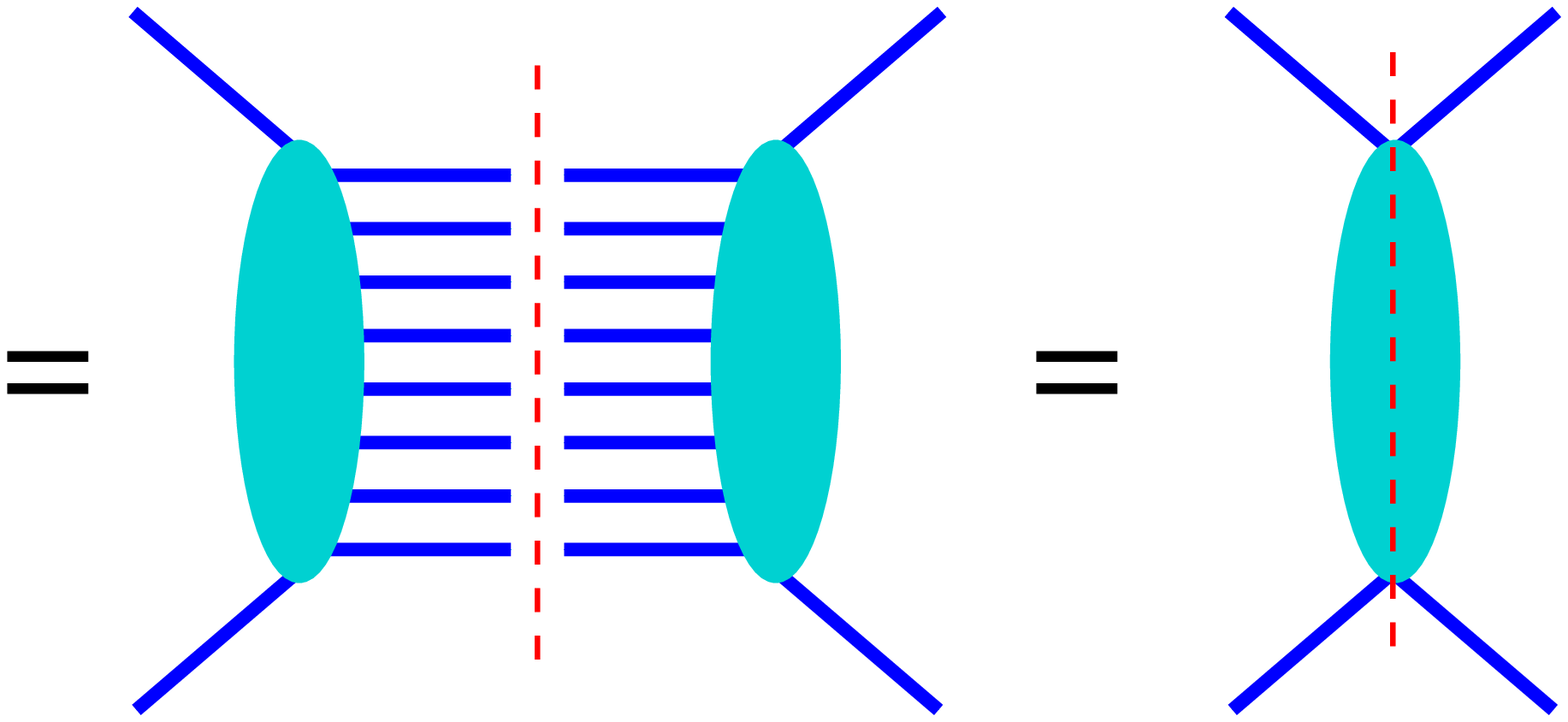}} \par}

\caption{The expression \protect\( \sum _{X}(T_{2\rightarrow X}).\protect \)\protect\( (T_{2\rightarrow X})^{*}\protect \)which
may be represented as a ``cut diagram''.\label{cut}}
\end{figure}

Let us first introduce some conventions. We denote elastic two body scattering
amplitudes as \( T_{2\rightarrow 2} \) and inelastic amplitudes corresponding
to the production of some final state \( X \) as \( T_{2\rightarrow X} \)
(see fig. \ref{t}). 
As a direct consequence of unitarity on may write the optical theorem  \( 2\mathrm{Im}T_{2\rightarrow 2}=\sum _{x} \)\( (T_{2\rightarrow X}) \)\( (T_{2\rightarrow X})^{*} \).
The right hand side of this equation may be literally presented as a ``cut
diagram'', where the diagram on one side of the cut is \( (T_{2\rightarrow X}) \)
and on the other side \( (T_{2\rightarrow X})^{*} \), as shown in fig. \ref{cut}. 
So the term ``cut diagram'' means nothing but the square of an inelastic amplitude,
summed over all final states, which is equal to twice the imaginary part of
the elastic amplitude.

\begin{figure}[htb]
{\par\centering \resizebox*{!}{0.15\textheight}{\includegraphics{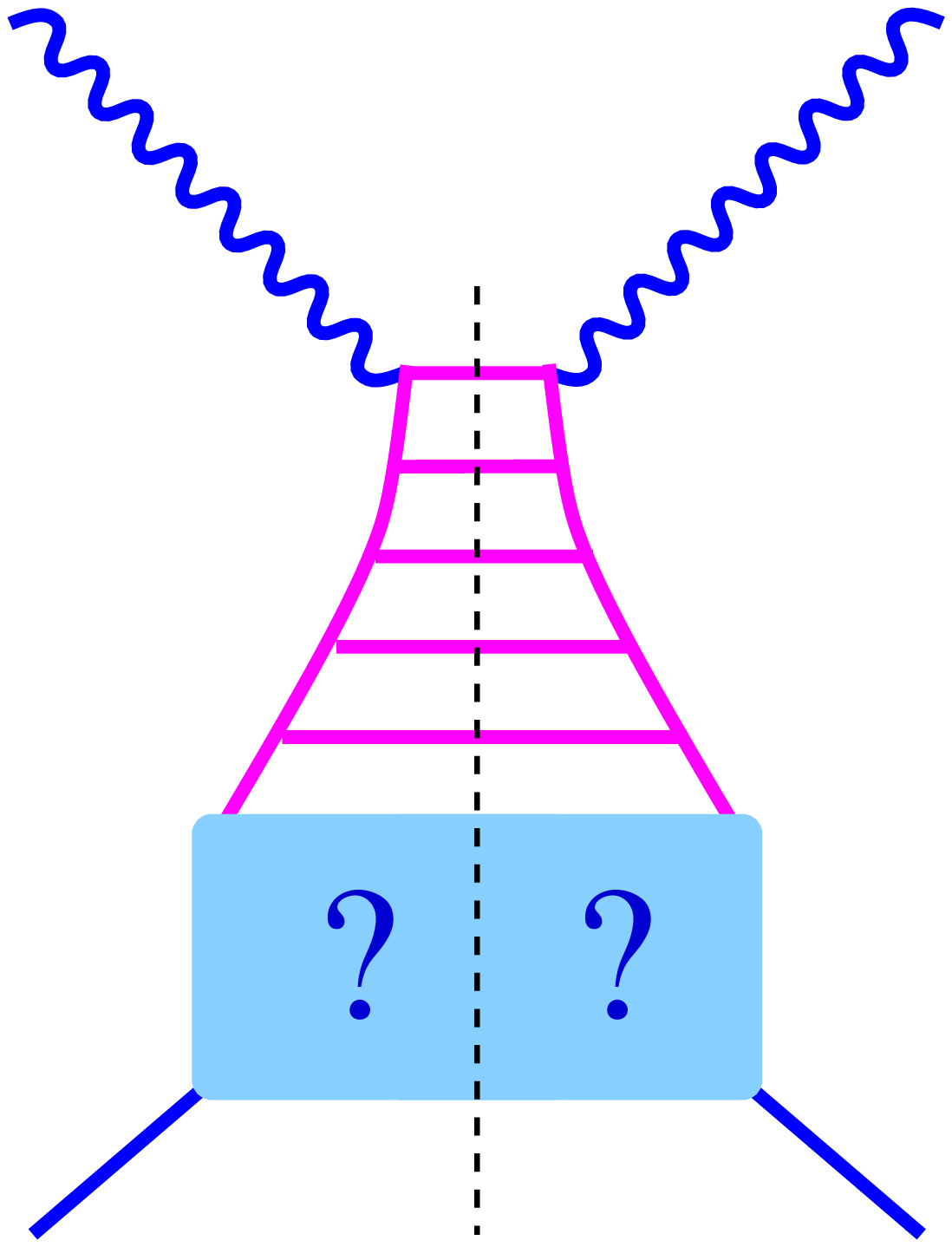}} 
\resizebox*{!}{0.15\textheight}{\includegraphics{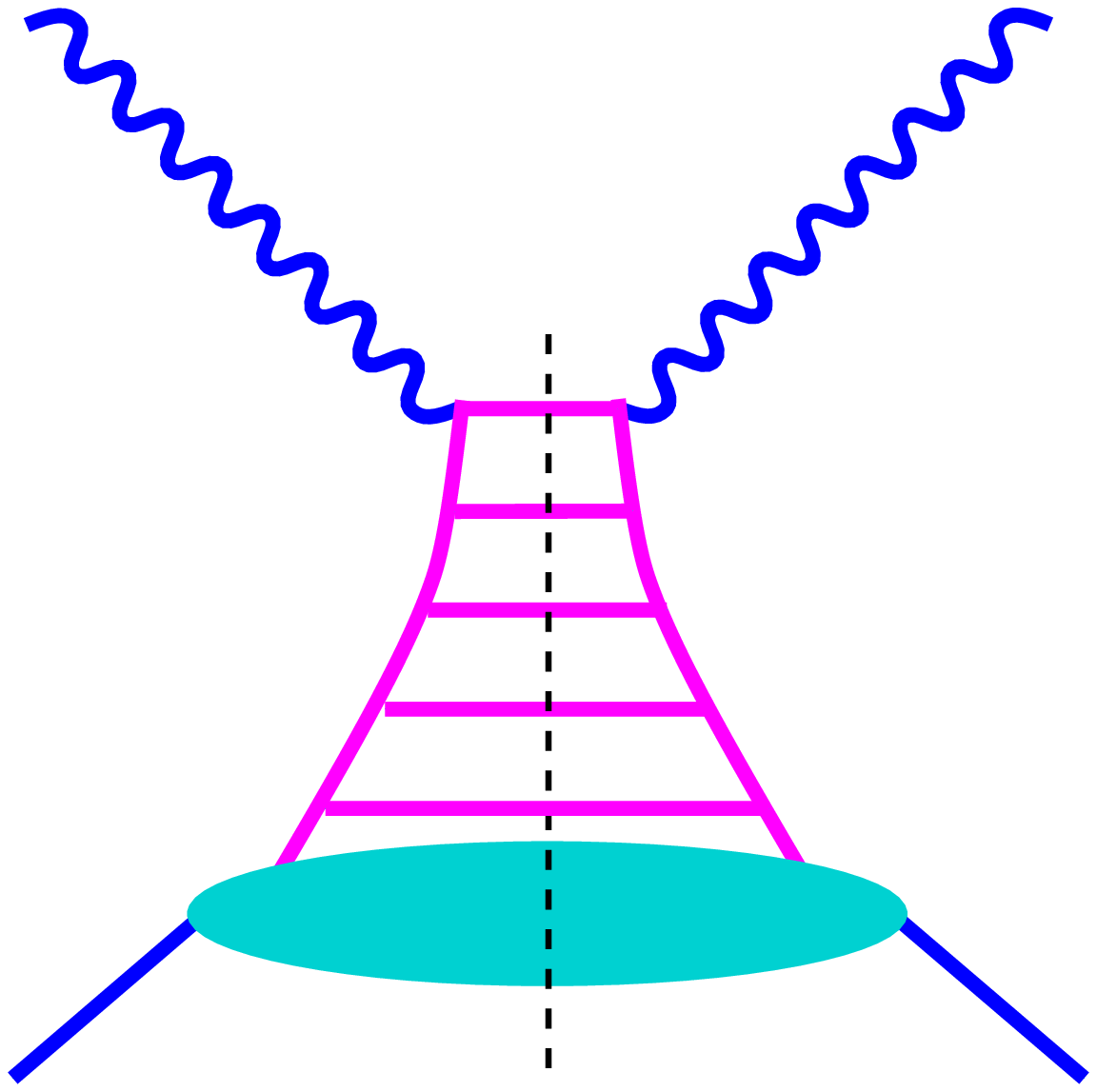}}  \resizebox*{!}{0.15\textheight}{\includegraphics{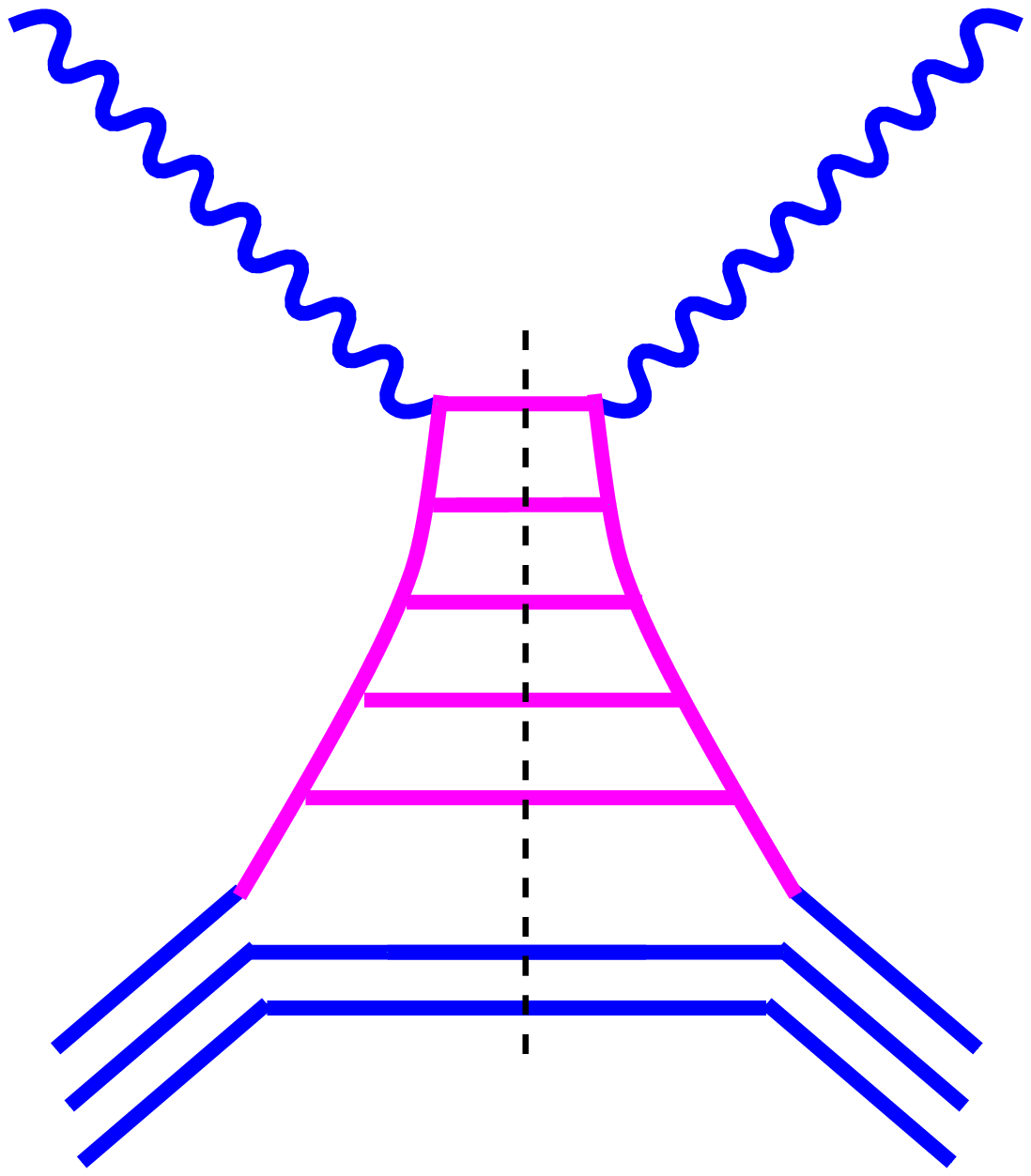}} \par}

\caption{Deep inelastic scattering: the general diagram (left) the ``sea'' contribution
with a soft Pomeron at the lower end (middle) and the ``valence'' contribution
(right).\label{dis}}
\end{figure}

Before coming to nuclear collisions, we need to discuss somewhat the structure
of the nucleon, which may be studied in deep inelastic scattering -- so essentially
the scattering of a virtual photon off a nucleon, see fig. \ref{dis}.
Let us assume a high virtuality photon. It is known that this photon couples
to a high virtuality quark, which is emitted from a parton with a smaller virtuality,
the latter on again being emitted from a parton with lower virtuality, and so
on. We have a sequence of partons with lower and lower virtualities, the closer
one gets to the proton. At some stage, some ``soft scale'' scale \( Q_{0}^{2} \)
must be reached, beyond which perturbative calculations are no longer valid.
So we have some ``unknown object'' -- indicated by ``?'' in fig. \ref{dis}
-- between the first parton and the nucleon. In order to proceed, one may estimate
the squared mass of this ``unknown object'', and one obtains doing simple
kinematics the value \( Q_{0}^{2}/x \), where \( x \) is the momentum fraction
of the first parton relative to the nucleon. Therefore, sea quarks or gluons,
having typically small \( x \), lead to large mass objects -- which we identify
with soft Pomerons, whereas valence quarks lead to small mass objects, which
we simply ignore. So we have two contributions, as shown in fig. \ref{dis}:
a sea contribution, where the sea quark or gluon is emitted from a soft Pomeron,
and a valence contribution, where the valence quark is one of the three quarks
of the nucleon. The precise microscopic structure of the soft Pomeron not being
known, it is parameterized in the usual way a a Regge pole.

\begin{figure}[htb]
{\par\centering \resizebox*{!}{0.11\textheight}{\includegraphics{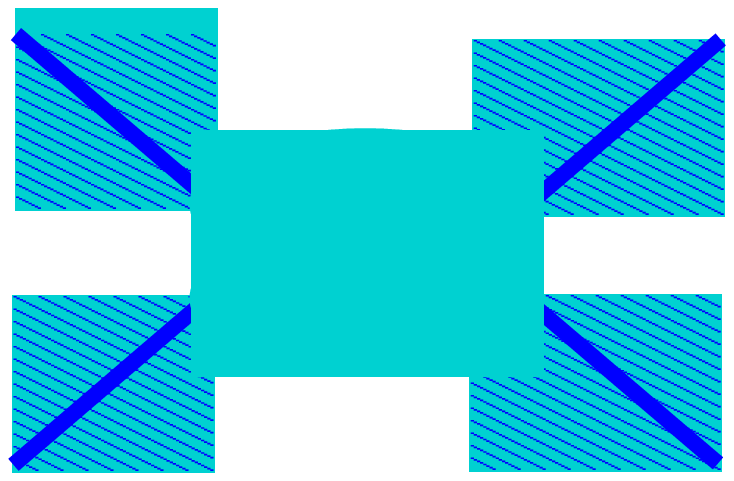}} \resizebox*{!}{0.15\textheight}{\includegraphics{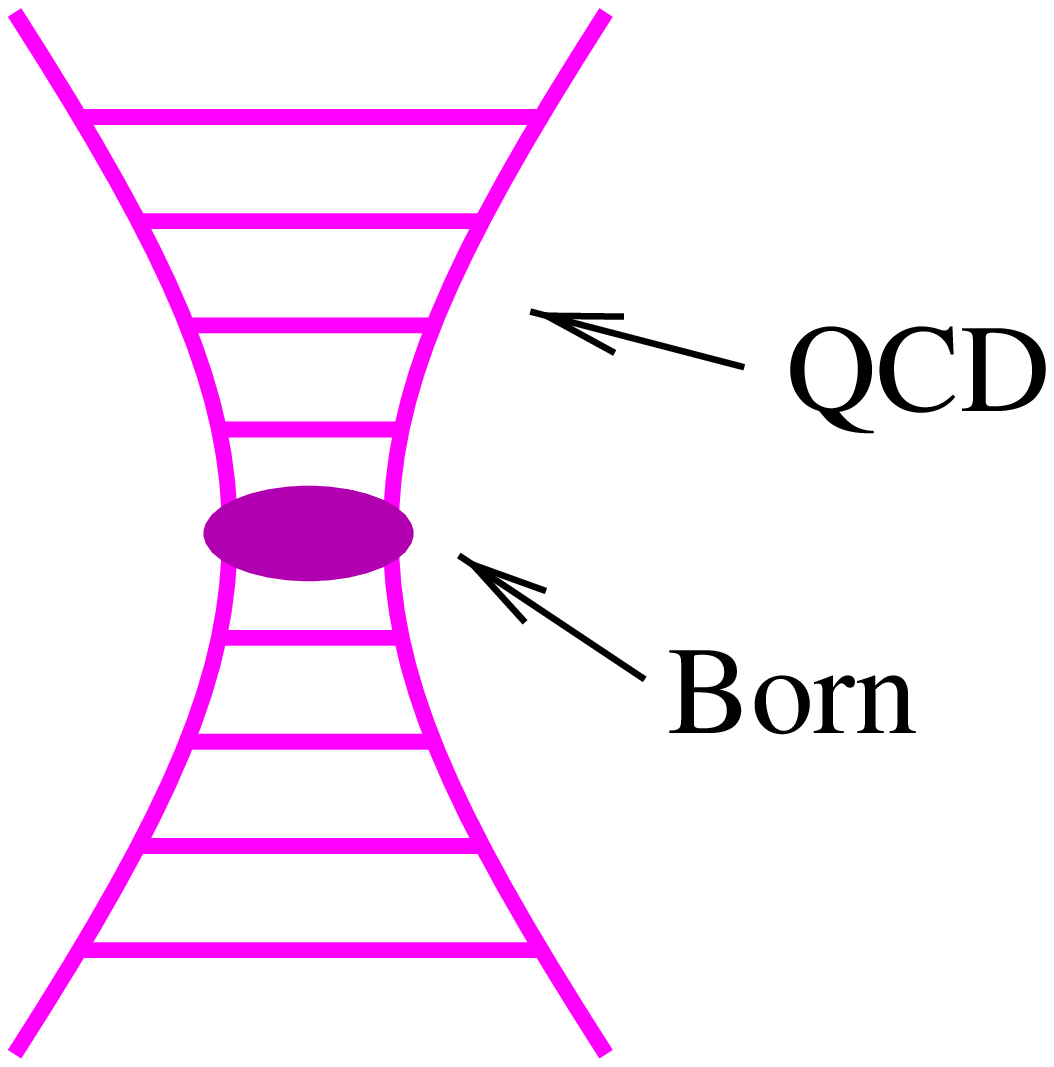}} \resizebox*{!}{0.15\textheight}{\includegraphics{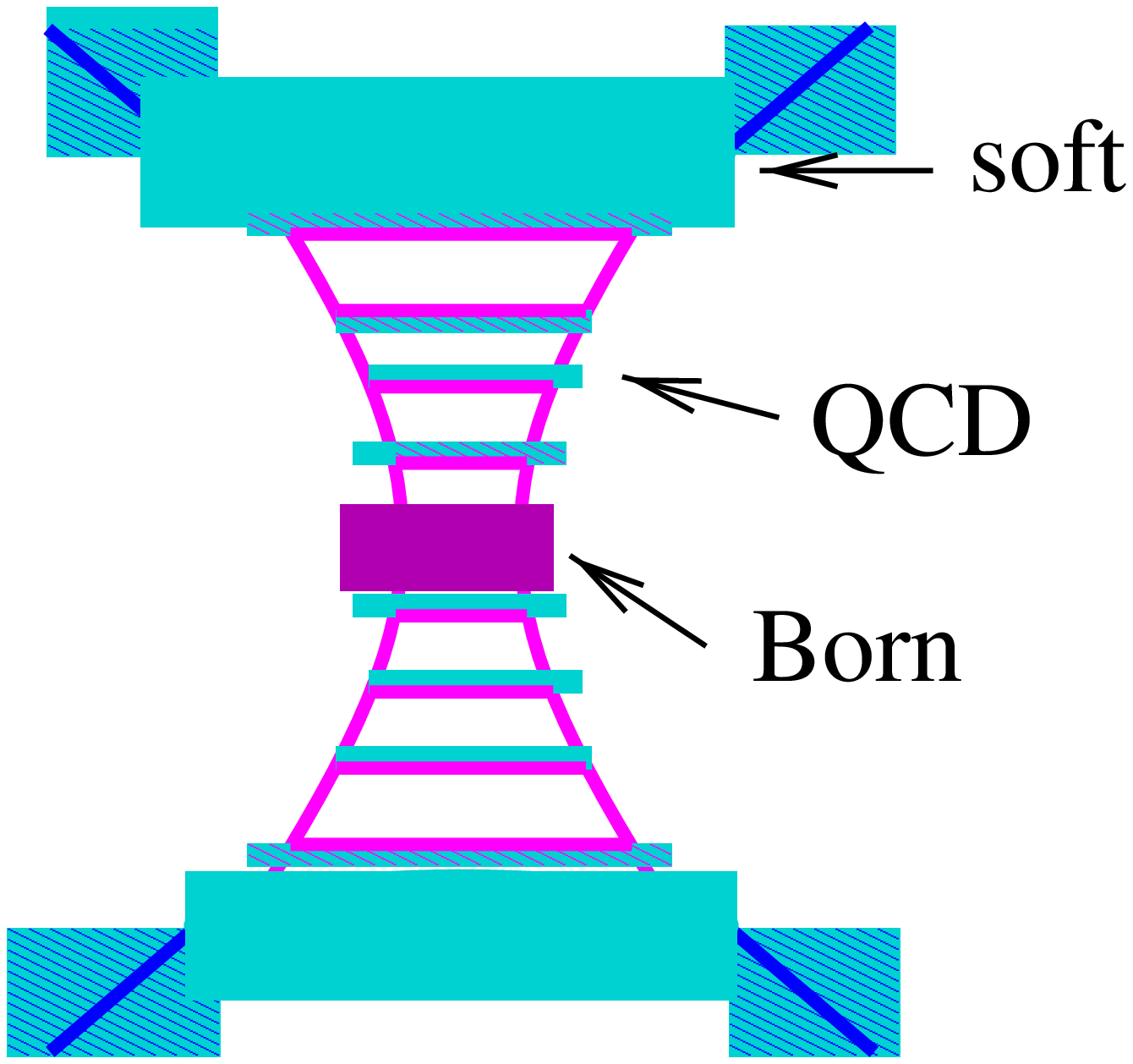}} \par}

\caption{The soft elastic scattering amplitude \protect\( T_{\mathrm{soft}}\protect \)
(left), the hard elastic scattering amplitude \protect\( T_{\mathrm{hard}}\protect \)
(middle) and one of the three contributions to the semi-hard elastic scattering
amplitude \protect\( T_{\mathrm{semi}}\protect \) (right). \label{nn}}
\end{figure}

Elementary nucleon-nucleon scattering can now be considered as a straightforward
generalization of photon-nucleon scattering: one has a hard parton-parton scattering
in the middle, and parton evolutions in both directions towards the nucleons.
We have a hard contribution \( T_{\mathrm{hard}} \) when the the first partons
on both sides are valence quarks, a semi-hard contribution \( T_{\mathrm{semi}} \)
when at least on one side there is a sea quark (being emitted from a soft Pomeron),
and finally we have a soft contribution, when there is no hard scattering at
all (see fig. \ref{nn}). 
We have a smooth transition from soft to hard physics: at low energies the soft
contribution dominates, at high energies the hard and semi-hard ones, at intermediate
energies (that is where experiments are performed presently) all contributions
are important.

Let us consider nucleus-nucleus (\( AB \)) scattering. In the Glauber-Gribov
approach \cite{gla59,gri69}, the nucleus-nucleus scattering amplitude is defined
by the sum of contributions of diagrams, corresponding to multiple elementary
scattering processes between parton constituents of projectile and target nucleons.
These elementary scatterings are exactly discussed above, namely the sum of
soft, semi-hard, and hard contributions: \( T_{2\rightarrow 2}=T_{\mathrm{soft}}+T_{\mathrm{semi}}+T_{\mathrm{hard}} \).
A corresponding relation holds for the inelastic amplitude \( T_{2\rightarrow X} \).
We use the above definition of a cut elementary diagram, which is graphically
represented by a vertical dashed line, whereas the elastic amplitude is represented
by an unbroken line:

\vspace{0.3cm}
{\par\centering \resizebox*{!}{0.05\textheight}{\includegraphics{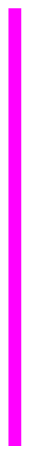}} \( \quad  \)\( =T_{2\rightarrow 2} \),
\( \quad  \)\( \quad  \)\resizebox*{!}{0.05\textheight}{\includegraphics{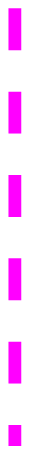}} \( \quad  \)\( =\sum _{X}(T_{2\rightarrow X}) \)\( (T_{2\rightarrow X})^{*} \).\par}
\vspace{0.3cm}

\begin{figure}[htb]
{\par\centering \resizebox*{!}{0.24\textheight}{\includegraphics{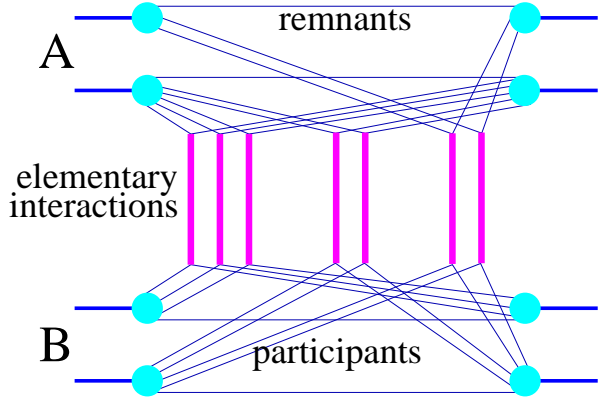}} \par}

\caption{Elastic nucleus-nucleus scattering amplitude, being composed of purely parallel
elementary interactions between partonic constituents of the nucleons (blobs).\label{aa}}
\end{figure}

\begin{figure}[htb]
{\par\centering \resizebox*{!}{0.24\textheight}{\includegraphics{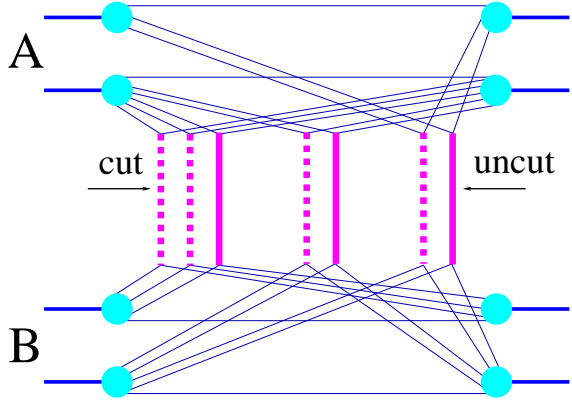}} \par}

\caption{Typical interference term contributing to the squared inelastic amplitude.\label{grtppaac}}
\end{figure}

\noindent This is very handy for treating the nuclear scattering model. We define
the model via the elastic scattering amplitude \( T_{AB\rightarrow AB} \) which
is assumed to consist of purely parallel elementary interactions between partonic
constituents, as discussed above. The amplitude is therefore a sum of terms
as the one shown in fig. \ref{aa}.
One has to be careful about energy conservation: all the partonic constituents
(lines) leaving a nucleon (blob) have to share the momentum of the nucleon.
So in the explicit formula one has an integration over momentum fractions of
the partons, taking care of momentum conservation. Having defined elastic scattering,
inelastic scattering and particle production is practically given, if one employs
a quantum mechanically self-consistent picture. Let us now consider inelastic
scattering: one has of course the same parallel structure, just some of the
elementary interactions may be inelastic, some elastic. The inelastic amplitude
being a sum over many terms -- \( T_{AB\rightarrow X}=\sum _{i}T^{(i)}_{AB\rightarrow X} \)
-- has to be squared and summed over final states in order to get the inelastic
cross section, which provides interference terms \( \sum _{X}(T^{(i)}_{AB\rightarrow X})(T^{(j)}_{AB\rightarrow X})^{*} \),
which can be conveniently expressed in terms of the cut and uncut elementary
diagrams, as shown in fig. \ref{grtppaac}. So we are doing nothing more than
following basic rules of quantum mechanics.
Of course a diagram with 3 inelastic elementary interactions does not interfere
with the one with 300, because the final states are different. So it makes sense
to define classes \( K \) of interference terms (cut diagrams) contributing
to the same final state, as all diagrams with a certain number of inelastic
interactions and with fixed momentum fractions of the corresponding partonic
constituents. One then sums over all terms within each class \( K \), and obtains
for the inelastic cross section

\noindent 
\[
\sigma _{AB}(s)=\int d^{2}b\: \sum _{K}\Omega ^{(s,b)}(K)\]
 where we use the symbolic notation \( d^{2}b=\int d^{2}b_{0}\, \int d^{2A}b_{A}\, \rho (b_{A})\, \int d^{2B}b_{B}\, \rho (b_{B}) \)
which means integration over impact parameter \( b_{0} \) and in addition averaging
over nuclear coordinates for projectile and target. The variable \( K \) is
characterized by \( AB \) numbers \( m_{k} \) representing the number of cut
elementary diagrams for each possible pair of nucleons and all the momentum
fractions \( x^{+} \) and \( x^{-} \) of all these elementary interactions
(so \( K \) is a partly discrete and partly continuous variable, and \( \sum  \)
is meant to represent \( \sum \int  \)). This is the really new and very important
feature of our approach: we keep explicitly the dependence on the longitudinal
momenta, assuring energy conservation at any level of our calculation. 

The calculation of \( \Omega  \) actually very difficult and technical, but
it can be done and we refer the interested reader to the literature \cite{dre00}. 

The quantity \( \Omega ^{(s,b)}(K) \) can now be interpreted as the probability
to produce a configuration \( K \) at given \( s \) and \( b \). So we have
a solid basis for applying Monte Carlo techniques: one generates configurations
\( K \) according to the probability distribution \( \Omega  \) and one may
then calculate mean values of observables by averaging Monte Carlo samples.
The problem is that \( \Omega  \) represents a very high dimensional probability
distribution, and it is not obvious how to deal with it. We decided to develop
powerful Markov chain techniques \cite{hla98} in order to avoid to introduce
additional approximations.

\section{Markov Chain Techniques}

In order to generate \( K \) according to the given distribution \( \Omega \left( K\right)  \),
defined earlier, we construct a Markov chain
\begin{equation}
\label{x}
K^{\left( 0\right) },K^{\left( 1\right) },K^{\left( 2\right) },...K^{\left( t_{\mathrm{max}}\right) }
\end{equation}
 such that the final configurations \( K^{\left( t_{\mathrm{max}}\right) } \)
are distributed according to the probability distribution \( \Omega \left( K\right)  \),
if possible for a \( t_{\mathrm{max}} \) not too large! To obtain a new configuration
\( K^{(t+1)}=L \) from a given configuration \( K^{(t)}=K \). We use Metropolis'
Ansatz for the transition probability \( p(K,L) \) as a product of a proposition
matrix \( w(K,L) \) and an acceptance matrix \( u(K,L) \). The detailed balance
condition -- which assures the convergence of the chain -- is automatically
fulfilled if \( u(K,L) \) is defined as 
\begin{equation}
\label{x}
u(K,L)=\min \left( \frac{\Omega (L)}{\Omega (K)}\frac{w(L,K)}{w(K,L)},1\right) .
\end{equation}
 We are free to choose \( w(K,L) \), but of course, for practical reasons,
we want to minimize the autocorrelation time, which requires a careful definition
of \( w \). An efficient procedure requires \( u(K,L) \) to be not too small
(to avoid too many rejections), so an ideal choice would be \( w\left( K,L\right) =\Omega \left( L\right)  \).
This is of course not possible, but we choose \( w(K,L) \) to be a ``reasonable''
approximation to \( \Omega (L) \) if \( K \) and \( L \) are reasonably close,
otherwise \( w \) should be zero. So we define 
\begin{equation}
\label{x}
w(K,L)=\left\{ \begin{array}{ccc}
\Omega _{0}(L) & \mathrm{if} & d(K,L)\leq 1\\
0 & \mathrm{otherwise} & 
\end{array}\right. ,
\end{equation}
 where \( d(K,L) \) is an integer quantity representing a distance between
two configurations (the maximum number of elementary interactions being different\cite{dre00}),
and where \( \Omega _{0} \) has a simple structure, just being a product of
terms \( \rho _{0} \) representing one single elementary interaction. So one
proposes only new configurations being ``close'' to the old ones. The above
definition of \( w(K,L) \) may be realized by the following algorithm:

\begin{itemize}
\item choose randomly a particular elementary interaction (say the \( \mu ^{\mathrm{th}} \)
interaction of the \( k^{\mathrm{th}} \) nucleon-nucleon pair)
\item propose a new configuration \( L \), which is obtained from the old one \( K \)
by removing the \( \mu ^{\mathrm{th}} \) interaction of the \( k^{\mathrm{th}} \)
nucleon-nucleon pair, and replacing this by a new one according to the distribution
\( \rho _{0} \).
\end{itemize}
This proposal is the accepted with a probability \( u(K,L) \). One should note
that proposing a configuration according to some ``approximation'' \( \Omega _{0} \)
of \( \Omega  \) is fully compensated by the acceptance procedure, so it is
an exact numerical solution of the problem, whatever be the precise definition
of \( \Omega _{0} \). 

This procedure works extremely well. We performed many test for situations where
conventional techniques work as well, and we find excellent agreement by using
\( 66.7\, k_{\mathrm{max}} \) iterations, where \( k_{\mathrm{max}} \) is
an upper limit estimate of the number of nucleon-nucleon interactions. The Markov-chain
method is perfect for our purposes, because we have fast convergence due to
the fact that \( \Omega _{0} \) is not too different from \( \Omega  \), on
the other hand one cannot use just \( \Omega _{0} \) to obtain an approximate
solution, because here we introduce an substantial error, which reaches for
example already on the nucleon-nucleon level about 100 \%.

\section{Summary}

We provide a new formulation of the multiple scattering mechanism in nucleus-nucleus
scattering, where the basic guideline is theoretical consistency. We avoid in
this way many of the problems encountered in present day models. We also introduce
the necessary numerical techniques to apply the formalism in order to perform
practical calculations.

This work has been funded in part by the IN2P3/CNRS (PICS 580) and the Russian
Foundation of Basic Researches (RFBR-98-02-22024). 

\bibliographystyle{pr2}
\bibliography{a}

\begin{thebibliography}{1}

\bibitem{gri68}
V.~N. Gribov,
\newblock Sov. Phys. JETP {\bf 26}, 414 (1968).

\bibitem{gri69}
V.~N. Gribov,
\newblock Sov. Phys. JETP {\bf 29}, 483 (1969).

\bibitem{dre00}
H.~J. Drescher, M.~Hladik, S.~Ostapchenko, T.~Pierog, and K.~Werner,
\newblock (2000), hep-ph/0007198,
\newblock to be published in Physics Reports.

\bibitem{bra90}
M.~Braun,
\newblock Yad. Fiz. (Rus) {\bf 52}, 257 (1990).

\bibitem{abr92}
V.~A. Abramovskii and G.~G. Leptoukh,
\newblock Sov.J.Nucl.Phys. {\bf 55}, 903 (1992).

\bibitem{sjo87}
T.~Sjostrand and M.~van Zijl,
\newblock Phys. Rev. {\bf D36}, 2019 (1987).

\bibitem{gla59}
R.~J. Glauber,
\newblock {\em in Lectures on theoretical physics} (N.Y.: Inter-science
  Publishers, 1959).

\bibitem{hla98}
M.~Hladik,
\newblock {\em Nouvelle approche pour la diffusion multiple dans les
  interactions noyau-noyau aux \'energies ultra-relativistes},
\newblock PhD thesis, Universit\'e de Nantes, 1998.

\end{thebibliography}

\end{document}